\documentclass{aa}  

\newcommand{\namesrc}{J2344}
\newcommand{\erasst}{eRASSt~J234402.9-352640}

\newcommand{\unitflux}{\text{erg\,cm}$^{-2}$\,\text{s}$^{-1}$}

\newcommand{\unitlumi}{\text{erg\,s}$^{-1}$}

\newcommand{\nswiftobs}{21}

\newcommand{\ndaysxraymonitoringapprox}{770}
\newcommand{\mjdnicerpresunblockstart}{59220}
\newcommand{\mjdnicerpresunblockstop}{59236}
\newcommand{\mjdnicerpostsunblockstart}{59332}
\newcommand{\mjdnicerpostsunblockstop}{59374}

\newcommand{\baldinimbhestimate}{$\log(M_{\mathrm{BH}}/\mathrm{M_{\odot}})=7.2 \pm 0.4$}

\newcommand{\erasstwolx}{$7.3 ^{+0.4}_{-0.4} \times 10^{44}$}
\newcommand{\absolutecmag}{-21.9}

\newcommand{\erasstwodbbktone}{$43^{+5}_{-4}$}
\newcommand{\erasstwodbbkttwo}{$113^{+8}_{-6}$}
\newcommand{\erassthreegamma}{$5.4^{+0.3}_{-0.2}$}
\newcommand{\erassfourgamma}{$6.3^{+1.0}_{-0.9}$}

\newcommand{\xmmonedbbktone}{$51^{+1}_{-2}$}
\newcommand{\xmmonedbbkttwo}{$106^{+3}_{-3}$}

\newcommand{\xmmtwoabspwrlawnh}{$19.97^{+0.08}_{-0.11}$}
\newcommand{\xmmtwoabspwrlawgamma}{$5.7^{+0.1}_{-0.1}$}

\newcommand{\xmmthreedbbktone}{$40^{+1}_{-1}$}
\newcommand{\xmmthreedbbkttwo}{$140^{+11}_{-9}$}

\newcommand{\tprec}{$3.0^{+0.6}_{-0.4}$}
\newcommand{\tprecfull}{$t_{\mathrm{p}}=3.0^{+0.6}_{-0.4}$}
\newcommand{\modulationfactorchange}{6}

\newcommand{\OBSMAXPOWER }{0.62}

\usepackage{graphicx}
\usepackage{txfonts}
\usepackage{natbib}
\usepackage{subcaption}
\usepackage{lscape}
\usepackage[dvipsnames]{xcolor}
\usepackage{orcidlink}

\begin{document}

   \title{Large-amplitude modulations and hours-timescale variability in the early X-ray light curve of a tidal disruption flare}

\author{A. Malyali\orcidlink{0000-0002-8851-4019}\inst{1} 
          \and
A.~Rau\orcidlink{0000-0001-5990-6243}\inst{1} \and
P.~Baldini\orcidlink{0009-0002-5164-3279}\inst{1} \and
A.~Franchini\orcidlink{0000-0002-8400-0969}\inst{2,3} \and
A.~G.~Markowitz\orcidlink{0000-0002-2173-0673}\inst{4} \and
A.~Merloni\orcidlink{0000-0002-0761-0130}\inst{1} \and
G.~E.~Anderson\orcidlink{0000-0001-6544-8007}\inst{5} \and
A.~J.~Goodwin\orcidlink{0000-0003-3441-8299}\inst{5} \and
D.~Homan\orcidlink{0000-0002-3243-874X}\inst{6} \and
M.~Krumpe\inst{6} \and
Z.~Liu\orcidlink{0000-0003-3014-8762}\inst{1} \and
J.~C.~A.~Miller-Jones\orcidlink{0000-0003-3124-2814}\inst{5} \and
I.~Grotova\orcidlink{0009-0007-3502-3412}\inst{1} \and
A.~Kawka\orcidlink{0000-0002-4485-6471}\inst{5}
          }
   \institute{Max-Planck-Institut f\"ur extraterrestrische Physik,  Giessenbachstrasse 1, 85748 Garching, Germany\\
              \email{amalyali@mpe.mpg.de}
         \and
Universit\`a degli Studi di Milano, Via Giovanni Celoria 16, 20134, Milano, Italy
\and
INFN, Sezione di Milano-Bicocca, Piazza della Scienza 3, I-20126 Milano, Italy
\and
Nicolaus Copernicus Astronomical Center, Polish Academy of Sciences, ul. Bartycka 18, 00-716 Warszawa, Poland
\and         
International Centre for Radio Astronomy Research - Curtin University, GPO Box U1987, Perth, WA 6845, Australia 
\and
Leibniz-Institut für Astrophysik Potsdam, An der Sternwarte 16, 14482 Potsdam, Germany
             }
   \date{Received NN, 2025; accepted NN, 2025}

% \abstract{}{}{}{}{} 
% 5 {} token are mandatory
 
  \abstract
  {
We present new X-ray, optical, and UV observations of the tidal disruption event candidate \erasst \,  (hereafter \namesrc). Between 50 and 60 days after peak optical brightness, \namesrc \, exhibited large-amplitude modulations in its 0.2--2~keV emission, when the flux repeatedly dimmed and re-brightened by a factor of $\sim$\modulationfactorchange \, over a $\sim$3-day timescale. These modulations exhibited harder-when-brighter behaviour but were not detected in high-cadence observations obtained 60--70 days and 170--200 days after peak optical brightness, when the system instead exhibited stochastic X-ray variability over timescales of hours. We discuss the different physical mechanisms responsible for such exotic X-ray variability and explore the possibility that the modulations in \namesrc \ were caused by the Lense-Thirring precession of the inner accretion flow around the disrupting black hole.
}


   \maketitle
%
%-------------------------------------------------------------------

\section{Introduction}\label{sec:intro}  
Stellar tidal disruption events (TDEs; \citealt{rees_tidal_1988}) offer a unique way to study accretion onto supermassive black holes (SMBHs) and probe their mass, $M_{\mathrm{BH}}$, and spin, $a$. In recent years, various efforts have explored constraining these properties from the emission produced by TDE flares. These include (i) fitting X-ray spectra using models of the TDE disc emission, i.e. non-stationary relativistic models of thin accretion discs (e.g. \citealt{mummery_spectral_2020}) and stationary slim accretion discs (\citealt{wen_continuum-fitting_2020,wen_library_2022}), (ii) fitting the multi-band optical-UV photometric light curves of TDEs using models that assume the optical emission is produced by stream-stream collisions (\citealt{ryu_measuring_2020}) or reprocessed accretion models (\citealt{guillochon_mosfit_2018,mockler_weighing_2019}), and (iii) scaling relationships between the late-time optical luminosities of TDEs and black hole masses \citep{mummery_fundamental_2023}.

This paper reports additional X-ray and photometric observations of the TDE candidate \erasst\ (hereafter \namesrc; \citealt{homan_discovery_2023}), which was discovered by the extended ROentgen Survey with an Imaging Telescope Array (eROSITA; \citealt{predehl_erosita_2021}) on board the Spektrum-Roentgen-Gamma (SRG) observatory (\citealt{sunyaev_srg_2021}) as an ultra-soft flare from a galaxy at $z=0.1$. The host galaxy contains a low-luminosity active galactic nucleus (\citealt{homan_discovery_2023}) with a central black hole mass of \baldinimbhestimate\ (Appendix~\ref{sec:appendix_bhmass}). In Sect.~\ref{sec:observations} we present an overview of the multi-wavelength follow-up campaign of \namesrc, before analysing the X-ray evolution in Sect.~\ref{sec:xray_evolution}. We then discuss the possible physical mechanisms driving the X-ray variability in Sect.~\ref{sec:discussion}, and summarise our work in Sect.~\ref{sec:summary}.
We adopted a flat $\Lambda$CDM cosmology throughout this work, with $H_0=67.7\, \mathrm{km}\, \mathrm{s}^{-1}\mathrm{Mpc}^{-1}$ and $\Omega _{\mathrm{m}}=0.309$ \citep{planck_collaboration_planck_2016}; $z=0.1$ therefore corresponds to a luminosity distance of 475~Mpc. All magnitudes are reported in the AB system, unless otherwise stated, and have been corrected for Galactic extinction using $A_{\mathrm{V}}=0.0386$\,mag \citep{schlafly_measuring_2011}, $R_{\mathrm{V}}=3.1,$ and a \citet{cardelli_relationship_1989} extinction curve. All dates and times are reported in universal time (UT), and we define the time of peak optical brightness, $t_{\mathrm{peak}}$, to be MJD=59170.

\section{Observations}\label{sec:observations}
\begin{figure}
    \centering        \includegraphics[scale=0.60]{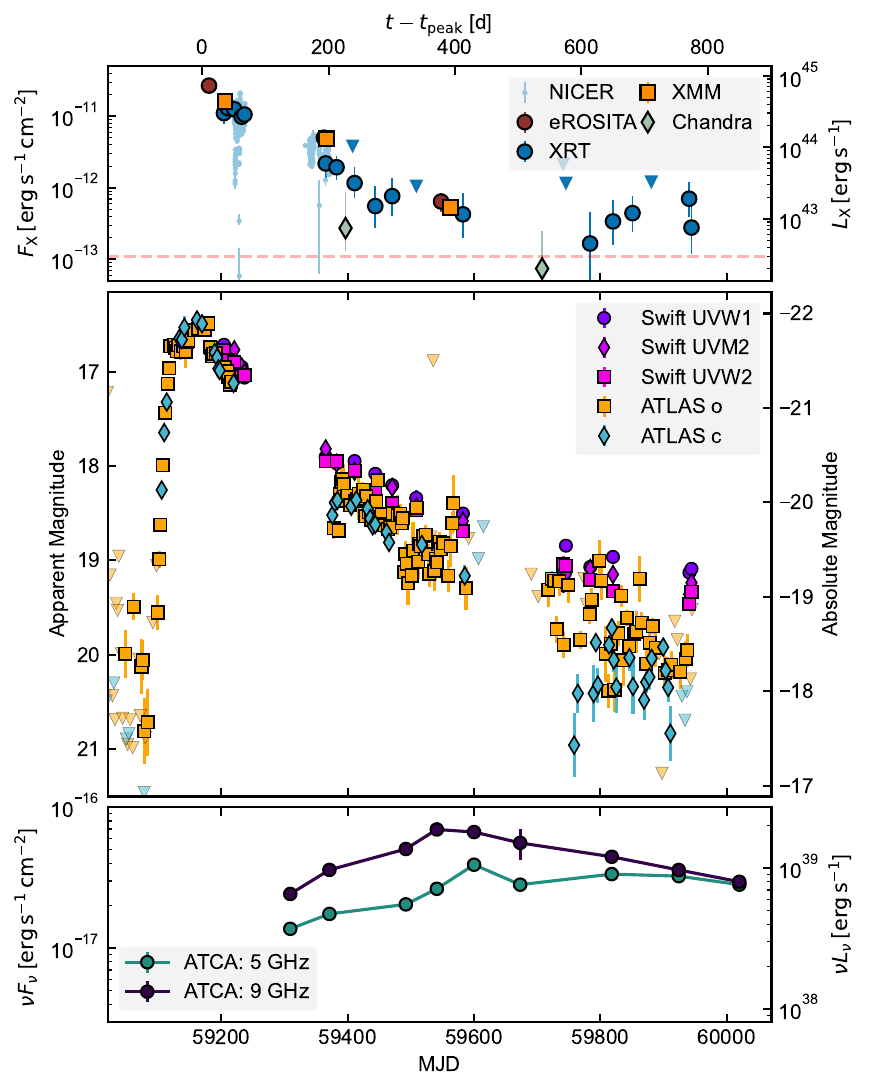} 
    \caption{Multi-wavelength light curve evolution of \namesrc , with the unabsorbed 0.2--2~keV fluxes (top panel) and the optical and UV fluxes (middle). Triangular data points denote 3$\sigma$ upper limits on the flux. The horizontal dashed red line in the X-ray light curve panel denotes the 3$\sigma$ upper limit on the 0.2--2~keV flux inferred from the non-detection of \namesrc ~ in eRASS1 \citep{homan_discovery_2023}, approximately 200 days before optical peak. The radio evolution (bottom panel) is well described by an expanding synchrotron-emitting region from a single ejection of material, consistent with an outflow launched by a non-relativistic TDE \citep{goodwin_radio_2024}.} 
    \label{fig:multiwavelength_evolution}
\end{figure}

A multi-wavelength follow-up campaign of \namesrc\, was initiated after its detection in the second eROSITA all-sky survey (eRASS2). Details regarding the observations and their associated data reduction are presented in Appendix~\ref{sec:appendix_data_reduction}. The extensive set of X-ray observations provided by eROSITA, \textit{Swift}/X-ray Telescope (XRT), XMM/EPIC-pn, the Neutron Star Interior Composition Explorer (NICER)/X-ray Timing Instrument (XTI), and \textit{Chandra}/Advanced CCD Imaging Spectrometer (ACIS)  makes \namesrc\, one of the best monitored non-jetted TDE candidates to date over the 0.2--10~keV range. The multi-wavelength evolution of \namesrc\  is plotted in Fig.~\ref{fig:multiwavelength_evolution}, and a summary of the X-ray flux evolution and optical/UV photometry is presented in Tables~\ref{tab:x_ray_lightcurve_table} and \ref{tab:uvot_photometry}, respectively.

% --------------------------

\begin{figure*}
    \centering
    \hfill
    \begin{subfigure}[t]{0.49\textwidth}
        \centering
        \includegraphics[scale=0.8]{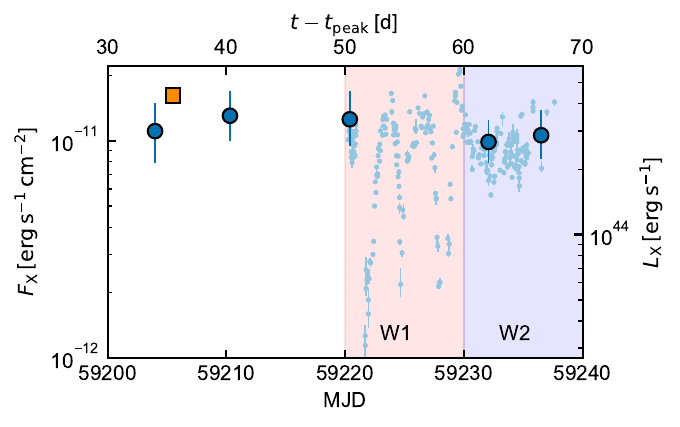} 
    \end{subfigure}
    \begin{subfigure}[t]{0.49\textwidth}
        \centering
        \includegraphics[scale=0.8]{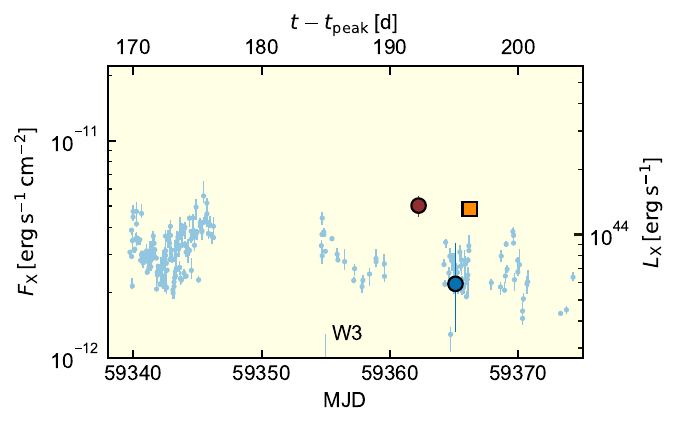} 
    \end{subfigure}
    \caption{Zoomed-in view of the X-ray evolution during the high-cadence NICER observations. The two panels have the same $y$-axis and cover the same amount of time on the $x$-axis. The different shaded backgrounds denote windows W1, W2, and W3 for the NICER observations (Sect.~\ref{sec:spec_nicer}), taken 50--60 days, 60--70 days, and 165--210 days after optical peak.
    } 
    \label{fig:nicer_evolution}
\end{figure*}

\section{X-ray evolution}\label{sec:xray_evolution}
\subsection{Spectral fitting}
The extracted X-ray spectra were fitted using the Bayesian X-ray Analysis (BXA) software (\citealt{buchner_x-ray_2014}), which connects the nested sampling algorithm UltraNest\footnote{\url{https://johannesbuchner.github.io/UltraNest/}} \citep{buchner_ultranest_2021}
with an X-ray spectral fitting environment. The eROSITA, XMM/EPIC-pn (\citealt{strueder_epicPN_2001}), \textit{Chandra}/ACIS, and \textit{Swift}/XRT spectra were fitted using CIAO/Sherpa \citep{fruscione_ciao_2006}, with their background spectra jointly modelled using a principle-component-analysis-based approach \citep{simmonds_xz_2018}. The NICER/XTI spectra were fitted using \texttt{XSPEC}, with the background contribution to the spectra modelled using the HEASOFT tool \texttt{SCORPEON}\footnote{\url{https://heasarc.gsfc.nasa.gov/docs/nicer/analysis_threads/scorpeon-overview/}}. The eROSITA, XMM/pn, \textit{Swift}/XRT, \textit{Chandra}/ACIS, and NICER/XTI spectra were fitted in the 0.2--8~keV, 0.2--10~keV, 0.5--7~keV, 0.3--10~keV, and 0.22--15~keV energy bands, respectively. All spectra were fitted using the C-statistic \citep{cash_generation_1976} and were unbinned for fitting. Lastly, each fitted spectral model was absorbed by a total ($\mathrm{HI}+\mathrm{H}_2$) Galactic hydrogen column density of $1.2 \times 10^{20}$~cm$^{-2}$ \citep{willingale_calibration_2013,homan_discovery_2023}, with abundances adopted from \citet{wilms_absorption_2000} and cross-sections from \citet{verner_atomic_1996}. 

\begin{figure}
    \centering
    \includegraphics[scale=0.82]{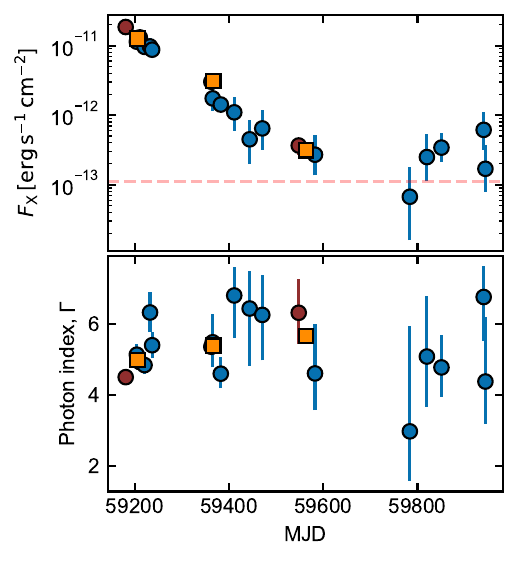}
    \caption{Joint evolution of the observed 0.2--2~keV flux (top panel) and the photon index (bottom panel) over time.  
    The dashed red line marks the 3$\sigma$ flux upper limit inferred from the non-detection in eRASS1.
    The X-ray spectrum remains ultra-soft ($\Gamma \gtrsim 4$) over the $\sim$760 days of X-ray monitoring ($\sim 800$ days after optical peak; \citealt{homan_discovery_2023}).}
    \label{fig:xray_photon_index_evolution}
\end{figure}

Before exploring more complex spectral models, we fitted a simple power-law model (\texttt{tbabs*powerlaw}, with $N_{\mathrm{H}}=1.2 \times 10^{20}$~cm$^{-2}$) to each X-ray spectrum to gain a first-order understanding of the X-ray spectral evolution. Over the $\sim$\ndaysxraymonitoringapprox-day X-ray monitoring campaign, the X-ray spectra remained ultra-soft (photon indices $\gtrsim 4$; Fig.~\ref{fig:xray_photon_index_evolution}) and without a clear active-galactic-nucleus-like power-law component produced by inverse-Compton-scattered disc photons (e.g. \citealt{nandra_ginga_1994}). The X-ray emission at late times is also in excess of the 3$\sigma$ pre-flare upper limit on the 0.2--2~keV flux of $1.1 \times 10^{-13}$~ \unitflux \citep{homan_discovery_2023} provided by the non-detection of J2344 in eRASS1, suggesting that this emission is associated with the recent enhanced accretion episode.

\subsubsection{eROSITA and XMM spectra}\label{sec:xray-erosita-xmm} 
The three eROSITA and three XMM spectra, obtained between $\sim$20 days and $\sim$400 days after the peak optical brightness, provide the strongest constraints on the X-ray spectral properties of J2344. We fitted eight different models (Table~\ref{tab:bxa_model_table}) to each spectrum, with the choice of models guided by previous models used in the literature to fit the ultra-soft spectra of TDEs, as well as the model fitting performed in \citet{homan_discovery_2023}. For each fitted model, we computed the Akaike information criterion, $AIC=2k - 2 \ln \hat{\mathcal{L}}$, where $k$ is the number of free parameters in the model and $\hat{\mathcal{L}}$ is the maximum likelihood of the fit. The best fitting model to each spectrum is the one with the lowest $AIC$. The best fitting models to the eROSITA and XMM spectra are presented in Figs.~\ref{fig:bxa_erosita_spectra} and \ref{fig:bxa_xmm_spectra}, respectively, and the spectral fit results are presented in Table~\ref{tab:x_ray_model_fits}.

The preferred model fit for the eRASS2 spectrum is the dual blackbody (\texttt{zbbody+zbbody}) model, with $kT_1=$\erasstwodbbktone \, eV and $kT_2=$ \erasstwodbbkttwo \, eV. For the fainter eRASS3 and eRASS4 spectra, a power-law model (\texttt{zpowerlaw})  is preferred, with $\Gamma=$\erassthreegamma \, and \erassfourgamma , respectively.
For XMM1 and XMM3, the preferred fit is again the  double blackbody model, with $kT_{1}=$\xmmonedbbktone ~eV and $kT_{2}=$\xmmonedbbkttwo ~eV in XMM1, and $kT_{1}=$\xmmthreedbbktone ~eV and $kT_{2}=$\xmmthreedbbkttwo ~eV in XMM3. However, the preferred fitted model for the XMM2 spectrum is an absorbed power law (\texttt{ztbabs*zpowerlaw}), with $\log [N_{\mathrm{H}} / \mathrm{cm}^2]=$\xmmtwoabspwrlawnh \, and $\Gamma=$\xmmtwoabspwrlawgamma, with the dual blackbody model under-predicting the model flux in the 1--2~keV range (Fig.~\ref{fig:bxa_xmm_spectra}).

\subsubsection{\textit{Chandra} and \textit{Swift} spectra}\label{sec:bxa_chandra_swift}
Since the \textit{Swift}/XRT and \textit{Chandra} spectra were fitted over the 0.3--10~keV and 0.5--7~keV ranges, respectively, and possess lower signal/noise values, these observations place weaker constraints on the spectral properties of the ultra-soft emission in J2344 in comparison to the eROSITA and XMM spectra. Therefore, we focused on using these spectra to measure 0.2--2~keV fluxes and complete J2344's X-ray light curve, leaving spectral parameter estimation to the higher signal/noise spectra (Sect.~\ref{sec:xray-erosita-xmm}). To do this, we fitted each spectrum with a dual blackbody model where $kT_{1}$ and $kT_2$ were constrained to between 35~eV and 55~eV, and 100~eV and 140~eV, respectively, whilst the normalisations of each component were left free to vary. For \textit{Swift}/XRT epochs where J2344 was not significantly detected above the background, the 0.2--2~keV flux upper limits were generated from the 0.3--2~keV count rate upper limits (Appendix~\ref{sec:dr_xrt}) using webPIMMs\footnote{\url{https://heasarc.gsfc.nasa.gov/cgi-bin/Tools/w3pimms/w3pimms.pl}}; we adopted the spectral model inferred from our fit to the first XMM spectrum to convert from count rates to fluxes.

\subsubsection{NICER spectra}\label{sec:spec_nicer}
Each NICER spectrum generated from a good time interval (GTI) with exposure above 100~s (Appendix~\ref{sec:dr_nicer}) was fitted with the same constrained phenomenological dual blackbody model used for the \textit{Swift}/XRT and \textit{Chandra} spectra (Sect.~\ref{sec:bxa_chandra_swift}). \citet{homan_discovery_2023} also fitted a dual blackbody model to the stacked NICER spectra over the 0.3--2~keV range using the first $\sim$6 days of observations, finding $kT_1=59 \pm 1$~eV and $kT_2=105 \pm 3$~eV; they also found  that single blackbody and single power-law models provided poor fits to these spectra.

\begin{figure}
    \centering
    \includegraphics[scale=0.8]{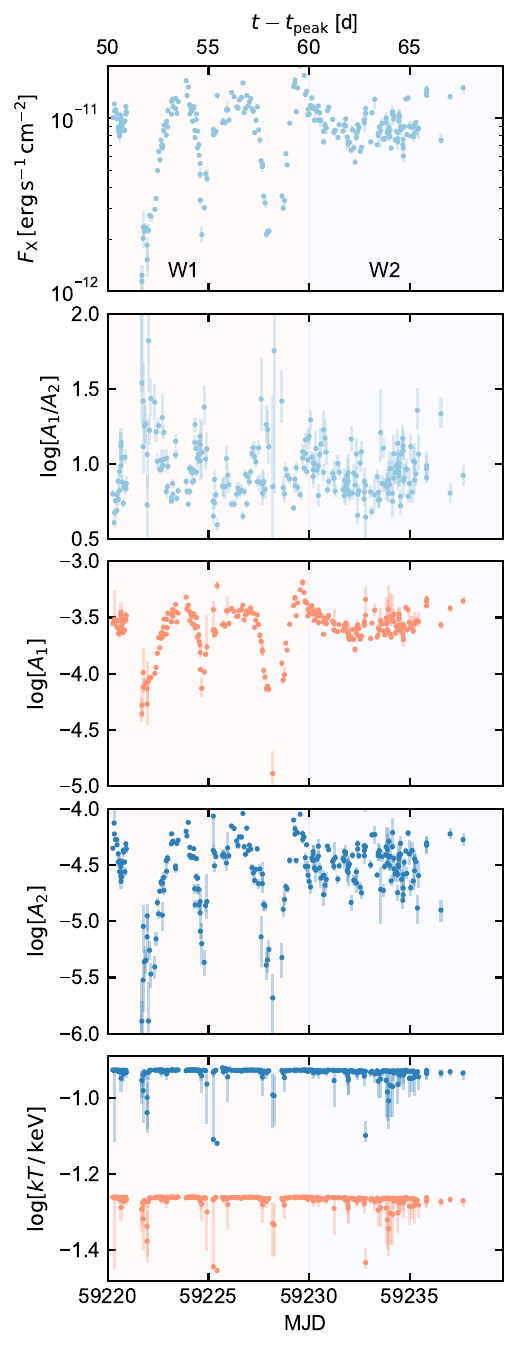}
    \caption{Top panel: Evolution of the 0.2--2~keV flux, $F_{\mathrm{X}}$, over time during the high-cadence NICER observations at early times ($\sim$50 days after optical peak). Second panel: spectral hardness estimated as $\log [A_1 / A_2]$, with  $A_1$ and $A_2$  the normalisation of the softer and harder blackbody components, respectively. Since smaller $\log [A_1 / A_2]$ values correspond to harder spectra, J2344 exhibits a harder-when-brighter behaviour. Third and fourth panels: log$A_1$ and log$A_2$ in units of $L_{39}/[D_{10}(1+z)]^2$, with $L_{39}$ the luminosity in $10^{39}$ erg~s$^{-1}$ and $D_{10}$ the distance to the source in units of 10~kpc. Bottom panel: Evolution of the blackbody temperatures for each model component. The $kT_1$ (orange) and $kT_2$ (blue) were constrained to the 35-55~eV and 100-140~eV ranges during fitting.}
    \label{fig:nicer_spectral_evolution}
\end{figure}

To aid in describing the spectral evolution during the NICER observations, we refer to NICER observations taken 50--60 days, 60--70 days, and 165--210 days after optical peak as windows 1, 2, and 3, respectively (i.e. W1, W2, and W3; Fig.~\ref{fig:multiwavelength_evolution}). 
In W1, the 0.2--2~keV flux undergoes a series of large-amplitude modulations during which the flux drops by a factor of $\sim$\modulationfactorchange \, over 1.5 days before re-brightening over the following 1.5\,days, with this cycle repeating two more times in W1 (a more detailed timing analysis is presented in Sect.~\ref{sec:xray_lightcurve_evolution}). During W1, the emission in the 0.2--2~keV band is dominated by the softer blackbody, the spectrum is harder when brighter (Figs.~\ref{fig:nicer_spectral_evolution} and~\ref{fig:nicer_spectral_evolution_simple}), and the normalisation of the two blackbody components increases when the system is in the brighter stage of a modulation. In W2 and W3, the modulations are not observed and the 0.2--2~keV flux evolution is more stochastic, with the emission dominated by the softer blackbody. We note that W2 immediately follows W1, implying that the modulations disappeared abruptly.

%-----------------
% --------------------------

\subsection{X-ray light curve evolution}\label{sec:xray_lightcurve_evolution}
\subsubsection{Long-term evolution}
\begin{figure}[h]
    \centering
    \includegraphics[scale=0.8]{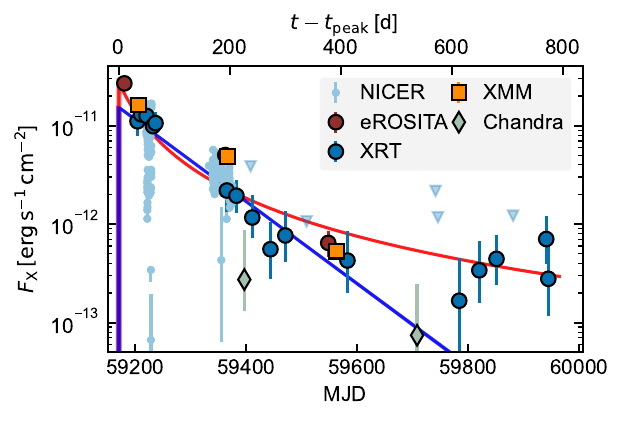}
    \caption{Power-law decay (red) and exponential decay (blue) models fitted to the 0.2--2 keV X-ray light curve of J2344 (excluding the early-time NICER data that exhibit modulations). The late-time flattening in the X-ray fluxes is better fitted by the power-law model here.}
\label{fig:xray_long_term_evolution}
\end{figure}

We fitted the 0.2--2 keV light curve with an exponential decline model, $F_{\mathrm{X}}(t)=F_{\mathrm{peak}} \exp \left[ -(t-t_{\mathrm{peak}})/ \tau \right]$,  and a power-law model, $F_{\mathrm{X}}(t)=F_{\mathrm{peak}}( 1 + (t-t_{\mathrm{peak}})/t_0)^{-p}$. Here, the NICER data taken before MJD~59240 were excluded when the X-ray light curve exhibited modulations. In addition, we constrained $t_{\mathrm{peak}}$ to 59165 to 59175 for both models (based on the observed optical peak; Fig.~\ref{fig:multiwavelength_evolution}), and fixed $p$ to $5/3$ for the power-law model.  
The inferred $F_{\mathrm{peak}}$ is $ 2.9^{+0.3}_{-0.3} \times 10^{-11}$~\unitflux\ for the power-law model with $t_0 = 55 ^{+3}_{-3}$~days, whilst $F_{\mathrm{peak}}=1.54^{+0.05}_{-0.05} \times 10^{-11}$~\unitflux and $\tau = 104.2 ^{+0.7}_{-0.7}$~days for the exponential decay. These fitted models are shown in Fig.~\ref{fig:xray_long_term_evolution}. Neither model fits the data particularly well: the exponential model under-predicts the peak flux and the late-time X-ray emission, while the power-law model over-predicts the flux at $\sim 200-600$~days after optical peak but fits the X-ray fluxes $>$600~days after optical peak reasonably well.

\subsubsection{Early-time modulations and hours-timescale X-ray variability}\label{sec:xray_qpms}
\begin{figure}
    \centering
    \includegraphics[scale=0.72]{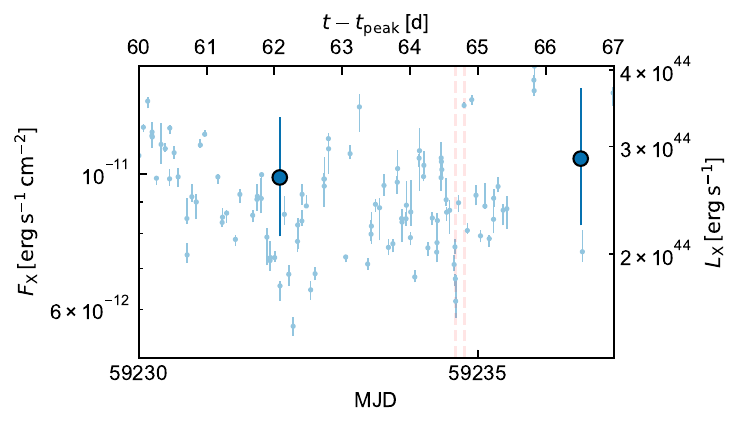}
    \caption{Zoom in on the hours-timescale stochastic X-ray variability during W2, after the shut-off of the large amplitude modulations. The two dashed red lines are separated by $\sim$3 hours, and the X-ray flux varies by a factor of 2 between these observations.}
    \label{fig:xray_plateau_variability_zoomed}
\end{figure}

The joint set of X-ray observations of \namesrc \,  reveals large-amplitude modulations in its 0.2--2 keV emission (Fig.~\ref{fig:multiwavelength_evolution}), with three clear modulations detected during the initial high-cadence NICER monitoring observations performed in W1 ($\sim50 - 60$ days after $t_{\mathrm{peak}}$). 
After the re-brightening observed in the last full modulation ($\sim$60 days after optical peak), the X-ray flux initially starts to decrease, similarly to the start of the other three modulations, but then begins exhibiting stochastic variations with a median flux of $F_{\mathrm{X, unabsorbed}}=8.9 \times 10^{-12}$ \, \unitflux, fainter than the peak flux observed during one of the modulations ($F_{\mathrm{X, unabsorbed}} \approx 1.3 \times 10^{-11}$ \, \unitflux). Modulations were also not detected in the post-sunblock NICER monitoring window at $\sim 170 - 200$ days after $t_{\mathrm{peak}}$, with the 0.2--2 keV source flux in these late-time observations almost an order of magnitude brighter than the sensitivity limits of NICER (i.e. we should have been able to detect similar-amplitude modulations in this monitoring window if they were still present). \namesrc \, also exhibits hours-timescale variability during W2 and W3, where the X-ray flux is observed to change by up to a factor of $\sim 2$ over $\sim$4-hour timescales\footnote{Similar X-ray variability behaviour was also observed during XMM1, when the X-ray flux varied by a factor of $\sim$2 over a three-hour timescale \citep{homan_discovery_2023}.} (Fig.~\ref{fig:xray_plateau_variability_zoomed}), similar to the hours-timescale X-ray variability reported in \citet{yao_subrelativistic_2024}. 

The early-time X-ray light curve also shows tentative evidence of periodic behavior. The   Lomb-Scargle periodogram (LSP; \citealt{lomb_least-squares_1976,scargle_studies_1982}) computed from the pre-sunblock light curve between MJD 59215 and 59240 reveals a clear peak at \tprec ~d, which we consider statistically significant at $> 4\sigma$ (see Appendix~\ref{sec:appendix_periodicity} for a detailed discussion). This periodicity is driven by the repeated modulations in W1 described above, since no significant peak is observed in the LSP when computed from NICER data taken only in W2. The phase-folded light curve in Fig.~\ref{fig:folded_lc} highlights that the X-ray spectrum softens in each modulation cycle as the flux drops; although the peak flux in each modulation is roughly constant, the shape of the dip is asymmetric, with the dimming faster than the brightening in each modulation. However, we caution that the major caveat to this detection of periodicity is that we have ultimately only observed three putative cycles, and pure red-noise processes can routinely mimic 3--4 cycle sinusoid-like periodicities \citep{vaughan_false_2016}. Robust detections of quasi-periodicity in accreting active galactic nuclei in cases where the light curves exhibit a low number of cycles -- that is, fully ruling out that the signal is red noise -- have not been convincingly achieved yet.

\begin{figure}
    \centering
    \includegraphics[scale=0.8]{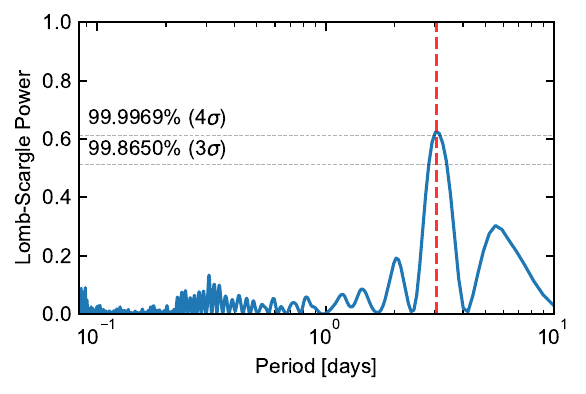}
    \caption{LSP computed from the 0.2--2 keV light curve between MJD 59215 and 59240. The dashed red line marks the period \tprecfull ~days, inferred from the peak with maximum power. The two dashed grey lines mark the 0.998650 ($\approx 3 \sigma$) and 0.999969 ($\approx 4 \sigma$) upper quantiles of the distribution of maximum Lomb-Scargle powers computed from the synthetic light curves (Fig.~\ref{fig:max_power_histo}).}
    \label{fig:lsp_observed}
\end{figure}

\begin{figure}
    \centering
    \includegraphics[scale=0.8]{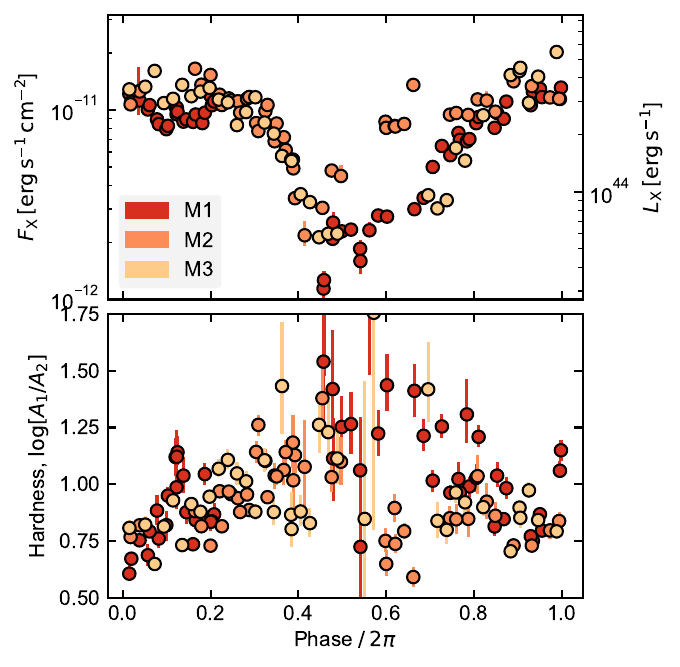}
    \caption{Folded 0.2--2~keV flux light curve (top panel) and hardness (bottom panel) of the \text{NICER} data from W1. In each modulation, the X-ray spectrum softens (higher $\log [A_1 / A_2]$) when the flux drops. The data points are coloured based upon which modulation they are associated with\ -- M1, M2, or M3 -- with M1 starting at $\mathrm{MJD}=59220.3$, and the light curves are folded with a periodicity of 3.05~days (this is consistent with the periodicity estimate of \tprec \, days provided by the LSP, but folding at 3.05~days provides a clearer presentation of each modulation than 3.0~days).}
    \label{fig:folded_lc}
\end{figure}

\section{Discussion}\label{sec:discussion}
In the following, we explore possible physical origins of the large-amplitude X-ray modulations observed in \namesrc. While we currently favour a precession-related scenario (see Sect.~\ref{sec:discussion_precession}), we also consider alternative explanations and outline the reasons they are less likely in Sect.~\ref{sec:disc_alternate_origins}.

\subsection{Precession of the inner accretion flow?}\label{sec:discussion_precession}
\begin{figure}
    \centering
    \includegraphics[scale=0.75]{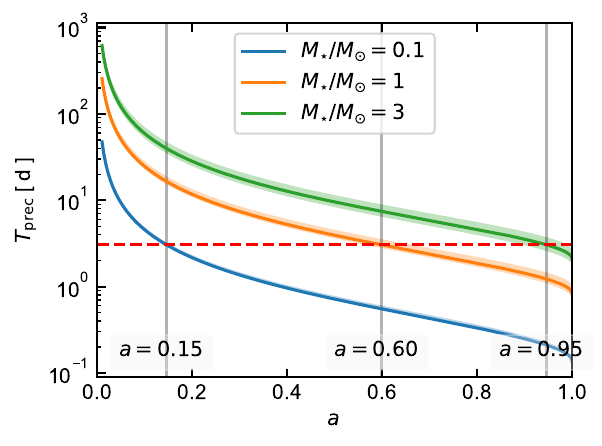}
    \includegraphics[scale=0.75]{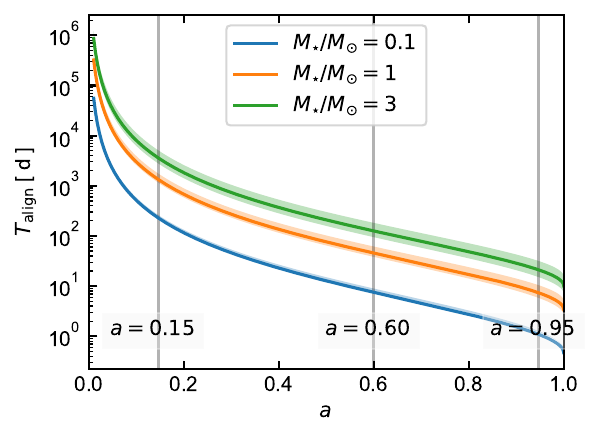}
    \caption{Precession (top panel) and alignment (bottom panel) timescales for a global precessing disc, plotted against black hole spin ($a$) for three stars of different masses being disrupted, assuming $\log [M_{\mathrm{BH}} / M_{\odot}]=7.2$ (Appendix~\ref{sec:appendix_bhmass}). The shaded regions enclose the uncertainty on each timescale due to the $1\sigma$ uncertainty on $\log [M_{\mathrm{BH}} / M_{\odot}]$. These timescales were computed with $H/R=0.1$ and $\alpha=0.1$, following \citet{franchini_lensethirring_2016}. 
    The dashed red line in the top panel denotes the observed 3-day modulation timescale observed in \namesrc, and the vertical grey lines denote the spin values that make $T_{\mathrm{prec}}$ consistent with 3~days (e.g. $a \sim 0.6$ for a Sun-like star), highlighting that spin constraints derived from this approach are sensitive to the assumed mass of the disrupted star. 
    }
    \label{fig:precession_spin_constraints}
\end{figure}

As tidally disrupted stars can approach the black hole isotropically \citep{stone_rates_2020}, a large fraction of TDEs will have misaligned stellar orbits and black hole equatorial planes. If the stellar debris circularises promptly around the black hole, then the newly formed accretion disc may be misaligned with the black hole's equatorial plane. However, since spinning black holes will exert a frame-dragging effect on the nascent accretion disc, a torque acts to align the angular momentum of the accretion disc with the black hole spin. For the thick, compact accretion discs predicted to form in TDEs,  the disc may initially precess as a solid body \citep{stone_observing_2012,shen_evolution_2014,franchini_lensethirring_2016}. 
As the disc precesses, a distant observer would periodically glimpse into the inner, hotter annuli of the disc, with the disc appearing brighter and harder when this occurs \citep{stone_stellar_2019}. The precession therefore leads to periodic modulations in the soft X-ray emission from the transient at early times, 
 as observed recently in AT~2020ocn \citep{pasham_lensethirring_2024}; this is also consistent with the X-ray light curve for \namesrc. Following \citet{franchini_lensethirring_2016} to estimate the precession period, and assuming the disruption of a Sun-like star by a black hole with $\log [M_{\mathrm{BH}}/ M_{\odot}]=7.2$, a spin of $a=0.6$ would yield a precession timescale consistent with the observed $3$-day timescale of modulations (Fig.~\ref{fig:precession_spin_constraints}). However, if the currently unknown mass of the disrupted star were lower (higher), then a lower (higher) black hole spin would be needed to produce a precession period consistent with the observed periodicity ($M_{\star} / M_{\odot}=0.1$ would need $a=0.15$, whilst $M_{\star} / M_{\odot}=3$ would need $a=0.95$; Fig.~\ref{fig:precession_spin_constraints}).
We note that the precession timescale also strongly depends on the disc surface density profile: a disc that is denser in the outer parts would precess on a much longer timescale. In the above, we assumed the surface density to be a decreasing function of radius, following the stabilisation of the radiation-pressure-dominated disc against the \cite{lightman_black_1974} instability, but a strong magnetic pressure can also stabilise the disc against the thermal instability and cause the surface density of the disc to decrease with radius \citep{kaur_magnetically_2023}.

At late times post-disruption, the X-ray modulations are predicted to `turn off' because the solid body precession cannot be maintained, either due to the disc thinning over time as the accretion rate decreases \citep{stone_observing_2012,shen_evolution_2014} or if the disc eventually becomes aligned with the black hole spin (see \citealt{franchini_lensethirring_2016} for a discussion of the relative importance of each effect in halting the solid body precession). Assuming an aspect ratio for the TDE disc of $H/R=1$ and disc viscosity parameter $\alpha = 0.1$, \citet{franchini_lensethirring_2016} predict an alignment timescale of $\sim 40$~days, which would be long enough for us to have observed the modulations in \namesrc \, at 50--60~days after optical peak. However, the sudden turn-off of the X-ray modulations observed in \namesrc \, between W1 and W2 (Fig.~\ref{fig:nicer_evolution}), which implies a sudden halting of the global disc precession, is different from the more gradual alignment behaviour predicted in \citet{franchini_lensethirring_2016}.

An alternative to the global disc precession scenario is that the X-ray modulations were caused by the precession of an inner disc after the tearing of the initial TDE disc into discrete rings \citep{nixon_tearing_2012}. For a disc broken at radius $R_{\mathrm{break}}\approx 16 R_{\mathrm{g}}(a/0.5)^{2/3}(\alpha / 0.1)^{-2/3}((H/R) / 0.1)^{-2/3}$ \citep{nixon_tearing_2012}, the inner disc material is expected to precess on the local Lense-Thirring timescale, $T_{\mathrm{prec}}\approx 4 (a/0.5)(\alpha / 0.1)^{-2}(H/R/0.1)^{-2}(M_{\mathrm{BH}}/ 10^{7.2} M_{\odot})$~days \citep{nixon_tearing_2012,raj_disk_2021}, and could therefore be capable of producing the $3$-day modulation timescale seen in \namesrc. As the inner ring spreads outwards over time, collisions may occur between neighbouring rings, leading to an increased accretion rate onto the black hole and a period of higher luminosity \citep{raj_disk_2021}. This would quench the modulations, similarly to what was observed in \namesrc \, between W1 and W2, and would be an advantage of using the disc tearing model to explain the modulations in \namesrc \, over the global precession model. We note, though, that the shape of the flares in the accretion rate from disc tearing differs from the dip-like modulations observed in \namesrc ~(see Fig.~3 in \citealt{raj_disk_2021}).

\subsection{Alternate origins}\label{sec:disc_alternate_origins}
The differential precession of the stellar debris streams and their violent self-intersection has been proposed to drive the early-time flaring in TDEs \citep{andalman_tidal_2021}, with periodic self-intersections driving periodic modulations of the accretion rate. However, they are predicted to only be present at early times during the disc formation process and therefore likely do not produce the modulations observed $\sim$50--60 days after the optical peak in \namesrc, since we expect the disc to already have been established by the time of the first eROSITA detection.

The harder-when-brighter behaviour of the system rules out the modulations being produced by lensing by a compact orbiting body, as has been considered to explain repeated X-ray outbursts in galactic centres \citep{ingram_self-lensing_2021}, and also rules out a variable neutral absorber. It is possible that this X-ray variability is linked to changing ionised absorption, but one would still need a mechanism to explain the repeated modulations every $\sim$3 days. A combination of a precessing disc and repeatedly viewing along a given angle of a disc wind, as seen in X-ray binaries (e.g. \citealt{kosec_vertical_2023}), may also explain the observed modulations. 

Sinusoidal-like modulations of the emission from a system involving a SMBH binary have been predicted. However, the system would be extremely close to merging if the modulation timescale of the X-rays in J2344 is associated with the SMBH binary's orbital period, and it would be highly unlikely to have observed such a system at $z=0.1$. Alternate variants of the binary model, such as the modulations being produced by collisions between an orbiting body and an accretion disc, are disfavoured due to the absence of such flares during NICER windows W2 and W3. Furthermore, such variants can require a large amount of fine-tuning to make the observed recurrence timescale of the outburst match the outburst duration.

Lastly, \citet{baldini_j2344_2026} report the detection of quasi-periodic eruptions from \namesrc\ in \textit{Einstein} Probe and \textit{XMM–Newton }observations obtained more than four years after the X-ray peak. These short-duration ($\sim$2~h) thermal flares recurred on a timescale of $\sim$12~h and persisted for at least nine months; they interpreted  them as arising from repeated interactions between a third body and the accretion disc formed during the TDE. Their markedly different phenomenology and characteristic timescales, relative to the modulations reported here, suggest that the two phenomena are physically distinct.

\section{Summary}\label{sec:summary}
The key observational signatures of \namesrc \, reported in this work are:
\begin{enumerate}
    \item An ultra-soft (photon indices $\gtrsim 4$), large-amplitude flare ($\gtrsim 150 \times$ brightening in the 0.2--2~keV band) from a low-luminosity active galactic nucleus, reaching a peak observed 0.2--2~keV luminosity of \erasstwolx ~\unitlumi. The X-ray emission remains ultra-soft over the course of the $\sim 800$ day follow-up campaign.
    \item \namesrc \, also shows transient optical and UV emission, with its optical light curve characterised by a fast rise to peak ($\sim 50$ days), followed by a decline period over at least the following $\sim 800$ days. Along with ASASSN-14li, \namesrc \, is one of the few TDE candidates that show a near contemporaneous optical and X-ray brightening, as well as a well-sampled, relatively smooth X-ray decline over months-long timescales.
    \item Between 50 and 60 days after optical peak, the high-cadence NICER observations reveal large-amplitude modulations of the ultra-soft X-ray emission, where the X-ray flux repeatedly dims and re-brightens by a factor of $\sim$6 over a $\sim3.1$-day timescale, and the X-ray emission becomes harder when brighter. These modulations are not detected in NICER observations 60--70 days and 170--200 days after the optical peak. 
\end{enumerate}
Although the physical mechanism producing these X-ray modulations is still unclear, we currently favour a precession-related origin (Sect.~\ref{sec:discussion_precession}) in order to explain the tentative 3-day periodicity in the early X-ray light curve.

\begin{acknowledgements}
We thank the anonymous referee for the insightful comments that helped improving the paper. AM is grateful to the generosity of Curtin University for hosting his visit, where parts of this work were completed. AM also thanks the Yukawa Institute for Theoretical Physics at Kyoto University, where discussions held during the YITP International Molecule-type Workshop YITP-T-19-07 on "Tidal Disruption Events: General Relativistic Transients" were useful for completing this work. AM thanks the \textit{Chandra}, NICER, \textit{Swift} and XMM teams for approving the ToO requests. AM acknowledges support by DLR under the grant 50 QR 2110 (XMM\_NuTra, PI: Z. Liu). This work was supported by the Australian government through the Australian Research Council’s Discovery Projects funding scheme (DP200102471). AGM acknowledges support from Narodowe Centrum Nauki (NCN) grant 2018/31/G/ST9/03224, and partial support from NCN grant 2019/35/B/ST9/03944.
This work is based on data from eROSITA, the soft X-ray instrument aboard SRG, a joint Russian-German science mission supported by the Russian Space Agency (Roskosmos), in the interests of the Russian Academy of Sciences represented by its Space Research Institute (IKI), and the Deutsches Zentrum für Luft- und Raumfahrt (DLR). The SRG spacecraft was built by Lavochkin Association (NPOL) and its subcontractors, and is operated by NPOL with support from the Max Planck Institute for Extraterrestrial Physics (MPE).
The development and construction of the eROSITA X-ray instrument was led by MPE, with contributions from the Dr. Karl Remeis Observatory Bamberg \& ECAP (FAU Erlangen-Nuernberg), the University of Hamburg Observatory, the Leibniz Institute for Astrophysics Potsdam (AIP), and the Institute for Astronomy and Astrophysics of the University of Tübingen, with the support of DLR and the Max Planck Society. The Argelander Institute for Astronomy of the University of Bonn and the Ludwig Maximilians Universität Munich also participated in the science preparation for eROSITA.
\end{acknowledgements}

\bibliographystyle{aa}
\bibliography{J234}

\begin{appendix}
\section{Black hole mass estimates}\label{sec:appendix_bhmass}
\citet{homan_discovery_2023} estimated the black hole mass of \namesrc \, with two methods based on optical spectroscopy, and provided an upper limit based on infrared photometry. The first estimate assumes that the black hole mass can be measured from the properties of the broad line region (BLR), using the Keplerian relation $M_{\mathrm{BH}}\propto RV^{2}$, where $R$ is the radial distance from the black hole and $V$ is the rotational velocity of the BLR. $V$ can be measured from the full width at half maximum (FWHM) of the broad lines, while reverberation mapping campaigns have found power-law relationships between $R$ and the continuum luminosity $L$ (e.g. \citealt{kaspi_relationship_2005}). The intrinsic scatter for this method is of the order of 0.4 dex, mainly due to uncorrelated variations between the continuum luminosity $L$ and the FWHM of the emission line \citet{shen_impact_2010}. 
In \citet{homan_discovery_2023}, both the $\rm H_{\mathrm{\beta}}$ and $\rm H_{\mathrm{\alpha}}$ lines were modelled with three Gaussian components: one narrow, one broad, and one very broad, with the broad and narrow components having the same centroid, while the very broad component is blue-shifted by 1000 km/s. The broad and very broad components were then averaged to derive an FWHM, which was used in the black hole mass estimation. However, the very broad components have been observed to be transient in additional follow-up spectroscopy (to be presented in a follow-up paper), and may instead be due to un-virialised, outflowing gas at early times post-disruption as in other TDE candidates (e.g. \citealt{nicholl_outflow_2020}), and for which the mass estimation technique described above is not applicable. 
Whilst the same might be true for the broad component, which is also transient, the absence of a significant blueshift makes it more likely that this is due to an illuminated virialised structure. Given the above considerations, we recompute the black hole mass from the FWHM of only the broad component of the line, obtaining the revised mass estimate of \baldinimbhestimate \, adopted in this work. It should be noted, though, that the applicability of this method to dynamical systems such as TDEs is not obvious. 

The second, less reliable mass estimate derived independently from the optical spectra is based on the $M_{\rm BH}-\sigma_*$ relation, connecting the black hole mass and the stellar velocity dispersion $\sigma_*$, with an intrinsic scatter of 0.44 dex (\citealt{gultekin_m_2009}). When $\sigma_*$ is not available, the line dispersion $\sigma_{\mathrm{[OIII]}}$ of the [OIII] 5007~$\AA$ line can be used by assuming that the dynamics of the narrow-line region gas are dominated by the potential of the host galaxy bulge (e.g., \citealt{salviander_black_2007}). Using the intrinsic scatter of the $\sigma_* - \sigma_{[OIII]}$ relation of 0.11 dex for sources in the redshift range of \namesrc, as reported by \citet{le_o_2023}, the value reported in \citet{homan_discovery_2023} of log($M_{\mathrm{BH}}/\mathrm{M_{\odot}})\sim7.9$ is compatible with our revised mass estimate of \baldinimbhestimate \, to within 2$\sigma$.

\section{Data reduction}\label{sec:appendix_data_reduction}

\subsection{eROSITA}\label{sec:dr_erosita}
The position of \namesrc \, was first observed by eROSITA in eRASS1 on 2020-05-(24-25), when no X-ray point source was detected within 60\arcsec of \namesrc \, and the inferred 3$\sigma$ upper limit on the observed 0.2--2~keV flux was $1.1 \times 10^{-13}$~ \unitflux \citep{homan_discovery_2023}. 
After the observed X-ray brightening of \namesrc\, detected by eROSITA in eRASS2 on 2020-11-(25-29), the system was later also observed in eRASS3 and eRASS4 on 2021-05-(27-28) and 2021-11-(29-30), respectively. 
Source and background spectra, and light curves, were generated using version 020 of the eROSITA event files and the eROSITA Science Analysis Software (eSASS\footnote{Version: eSASSusers\_211214.}; \citealt{brunner_erosita_2022}) pipeline task \texttt{SRCTOOL}. To do this, the source counts were extracted from a circular aperture of radius 30\arcsec, while the background counts were extracted from an annulus with inner and outer radii of 150\arcsec and 340\arcsec, respectively. Both apertures were centred on the \textit{Gaia} position of the optical transient associated with \namesrc.

\subsection{\textit{Swift}/XRT}\label{sec:dr_xrt}
J2344 was observed \nswiftobs\, times with the \textit{Swift}/X-ray Telescope (XRT; \citealt{burrows_xrt_2005}). Observations were performed in photon counting mode, with the early and late time observations performed on a near-weekly and monthly basis, respectively (excluding epochs of Sun block). We first generated a 0.3--2~keV count rate light curve for these observations through the online tool provided by the UK Swift Science Data Centre (UKSSDC; \citealt{evans_online_2007,evans_methods_2009}).

Then, XRT data were analysed with version 6.31 of the HEASoft analysis software, with XRT CALDB v20220803. Source and background spectra were generated using the task \texttt{xrtproducts}, with source counts extracted from a circular aperture of 47" radius centred on J2344, with background counts extracted from an annulus with inner and outer radii 70" and 250",
respectively. 

\subsection{\textit{XMM-Newton}}\label{sec:dr_xmm}
\namesrc\, was observed three times with \textit{XMM-Newton} on 22 December 2020 (PI: M. Krumpe), 2021-06-01 and 2021-12-14 (PI: Z. Liu), with observations obtained in imaging mode. The XMM data were reduced using Science Analysis Software (SAS) version 20211130\_0941, and the latest calibration data files. We first generated calibrated event files from the observation data files using the tasks \texttt{emproc} and \texttt{epproc} for the EPIC-MOS (\citealt{turner_epicMOS_2001}) and EPIC-pn (\citealt{strueder_epicPN_2001}) cameras, before filtering out periods of high background following the SAS recommended procedures\footnote{\url{https://www.cosmos.esa.int/web/xmm-newton/sas-thread-epic-filterbackground}}. For the first XMM observation, which was taken in small window mode and with the medium filter, only events with \texttt{PATTERN}==0 and \texttt{FLAG}==0 were extracted for the PN spectrum, as performed in \citet{homan_discovery_2023}. For the second and third XMM observations taken with the `thin1' filter, only events with \texttt{PATTERN}<=4 and \texttt{FLAG}==0 were extracted for PN. Using \texttt{evselect}, source spectra were extracted from a circle of radius 30 \arcsec, centred on the \textit{Gaia} optical position, whilst background spectra were extracted from a source-free circle of radius 30 \arcsec .

\subsection{NICER}\label{sec:dr_nicer}
Additional high-cadence observations of \namesrc\, were obtained using the XTI on board NICER (\citealt{den_herder_neutron_2016}; PIs: M. Krumpe, A. Malyali). These began on 6 January 2021 (MJD \mjdnicerpresunblockstart ), and continued multiple times per day for the following 16 days until 22 January 2021 (MJD \mjdnicerpresunblockstop ). 
After a pause of $\sim$100 days when J2344 was in sunblock, NICER observations then continued over the $\sim$40 days between 2021-04-28 (MJD \mjdnicerpostsunblockstart ) and 2021-06-09 (MJD \mjdnicerpostsunblockstop ). The NICER coverage during this second window features a mix of high-cadence observations (multiple per day), as well as several day gaps when no observations of J2344 were obtained (see Fig.~\ref{fig:multiwavelength_evolution}). In addition to presenting all new observations of J2344 taken after it first went into sunblock post-outburst (MJD$\gtrsim$59240), this work also presents an additional $\sim$12~days of high-cadence NICER observations of J2344 at early times that were not previously reported on in \citet{homan_discovery_2023}; this was made possible by the most recent enhancements to the NICER analysis software (HEASoft version 6.33.2) that has improved our modelling of X-ray spectra under high optical loading conditions. 

To generate cleaned and screened event files, we first ran the \texttt{nicerl2} task on the raw event file of each observation ID (ObsID). Data from noisy focal plane modules were excluded using the parameter \texttt{detlist=launch,-14,-34}, and we aimed to reduce the impact of optical loading by restricting the undershoot range using \texttt{underonly\_range=0-400}. This resulted in a cleaned event file and a set of GTIs associated with it, and we then generated spectral products for each GTI within an ObsID using the \texttt{nicerl3-spect} tool. No X-ray point sources were detected by eROSITA within a circle of area $\sim$30 arcmin${^2}$ (NICER's field of view) centred on the optical position of J2344, suggesting minimal contamination of the NICER source spectra by field X-ray point sources near to J2344.

\subsection{\textit{Chandra}}\label{sec:dr_chandra}
We obtained two observations (PI: A. Rau\footnote{\textit{Chandra} proposal ID: 22700358.}) of J2344 with the \textit{Chandra} ACIS (\citealt{garmire_advanced_2003}) on 2021-07-01 and 2022-05-08. 
We first reprocessed the \textit{Chandra} data products using \texttt{chandra\_reproc}, using CIAO version 4.14 and CALDB version 4.9.7. Then, source and background spectra were generated using the CIAO tool \texttt{specextract}, with \texttt{mkrmf} and \texttt{mkarf} being used for the generation of Redistribution Matrix Files (RMF) and Ancillary Response Files (ARF), respectively. Source counts were extracted from a circular region of 4.1\arcsec \, radius for both observations, while background counts were extracted from an annulus with inner radius of 9.1\arcsec \, for both spectra, and 46.4\arcsec \, and 60\arcsec \, for the first and second observations, respectively.

\subsection{Optical}\label{sec:optical_photometry}
The optical transient emission associated with J2344 was first publicly identified by the \textit{Gaia} alerts team\footnote{\url{http://gsaweb.ast.cam.ac.uk/alerts/alert/Gaia20eub/}}, which was later reported to the Transient Name Server (TNS) on 13 October 2020 as AT~2020wjw/ Gaia20eub\footnote{\url{https://www.wis-tns.org/object/2020wjw}}. Additional optical monitoring of J2344 has also been performed by the ATLAS \citep{tonry_atlas_2018} and ASAS-SN \citep{kochanek_asassn_2017} surveys. 
As we are interested in the optical variability of the nucleus of the host galaxy of J2344, and the publicly available \textit{Gaia} $G$-band photometry has not been generated from difference images, then we only use the ATLAS $o$ and $c$-band photometry (Table~\ref{tab:uvot_photometry}) in this paper. This was generated using the ATLAS online forced photometry tool \citep{smith_design_2020,shingles_release_2021} and ran on the \textit{Gaia} optical position reported for Gaia20eub. For MJD$<$59690 ($>$59690), we performed a weighted re-binning of the $o$ and $c$-band photometry into 1~day (5~day) bins. We note that the ATLAS photometry is also the deepest available and is used in lieu of the ASAS-SN photometry, presented in \citet{homan_discovery_2023}.

\subsection{\textit{Swift}/UVOT}
J2344 was observed with the \textit{Swift} Ultraviolet/Optical Telescope (UVOT; \citealt{roming_swift_2005}) in the $U$, $UVW1$, $UVM2$, $UVW2$ filters, with the same cadence as the XRT observations (Appendix~\ref{sec:dr_xrt}). The level 2 UVOT sky images were downloaded from the UK Swift Science Data Centre, and were analysed using version 6.31 of the HEASoft analysis software, with CALDB v20211108. Aperture photometry (Table~\ref{tab:uvot_photometry}) was then performed on these images with the \texttt{uvotsource} task, extracting source counts from a circular aperture of radius 5\arcsec \, centred on the \textit{Gaia} optical position, and background counts from an aperture of radius 15\arcsec \, centred on a source-free nearby background region. 

\section{Light curve datasets}
The X-ray light curve data is presented in Table~\ref{tab:x_ray_lightcurve_table}, whilst the optical and UV photometry can be found in Table~\ref{tab:uvot_photometry}. Complete tables are available in electronic format.

\begin{table*}
\centering
\caption{X-ray light curve of \namesrc.}
\label{tab:x_ray_lightcurve_table}
\begin{tabular}{cccccccc}
\hline
MJD & MJD$_{\mathrm{start}}$ & MJD$_{\mathrm{stop}}$ & Instrument & ObsID & $\log [F_{\rm 0.2-2keV, obs}]$  & $\log [F_{\rm 0.2-2keV, unabs}$] & $\log [L_{\rm 0.2-2keV}$] \\
\hline
59180.371 & 59178.611 & 59182.132 & eROSITA & eRASS2 & $-10.75 ^{+0.02}_{-0.02}$ & $-10.57 ^{+0.02}_{-0.02}$ & $44.86 ^{+0.02}_{-0.02}$ \\
59203.988 & 59203.983 & 59203.994 & XRT & 00013946001 & $-11.13 ^{+0.11}_{-0.12}$ & $-10.96 ^{+0.13}_{-0.15}$ & $44.48 ^{+0.13}_{-0.15}$ \\
59205.510 & 59205.440 & 59205.581 & EPIC pn & 0862770101 & $-10.97 ^{+0.00}_{-0.00}$ & $-10.79 ^{+0.01}_{-0.01}$ & $44.64 ^{+0.01}_{-0.01}$ \\
... & ... & ... & ... & ... & ... & ... & ...  \\
59940.465 & 59940.006 & 59940.924 & XRT & 00013946021 & $-12.36 ^{+0.21}_{-0.24}$ & $-12.15 ^{+0.23}_{-0.25}$ & $43.28 ^{+0.23}_{-0.25}$ \\
59944.203 & 59944.161 & 59944.245 & XRT & 00013946022 & $-12.74 ^{+0.31}_{-0.34}$ & $-12.56 ^{+0.34}_{-0.38}$ & $42.88 ^{+0.34}_{-0.38}$ \\
\end{tabular}
\tablefoot{
$F_{\rm 0.2-2keV, obs}$ and $F_{\rm 0.2-2keV, unabs}$ are the observed and unabsorbed 0.2--2~keV band fluxes in units of \unitflux.
}
\end{table*}

\begin{table}
\centering
\caption{Optical and UV photometry of \namesrc.}
\label{tab:uvot_photometry}
\begin{tabular}{cccc}
\hline
MJD & Instrument & Filter & Magnitude \\
\hline
57303.390 & ATLAS & o & $<$19.78 \\
57321.350 & ATLAS & o & $<$18.59 \\
57329.330 & ATLAS & o & $<$19.29 \\
... & ... & ... & ... \\
59944.210 & \textit{Swift} & UVW1 & 19.09 $\pm$ 0.06 \\
59945.110 & ATLAS & o & $<$19.52 \\
\end{tabular}
\tablefoot{
Photometry corrected for Galactic reddening. 3$\sigma$ upper limits are presented in cases where no significant detection was obtained.
}
\end{table}

\section{Further X-ray properties}\label{sec:appendix_xray_properties}

\subsection{X-ray spectral evolution}
\begin{figure}[h]
    \centering
    \includegraphics[scale=0.8]{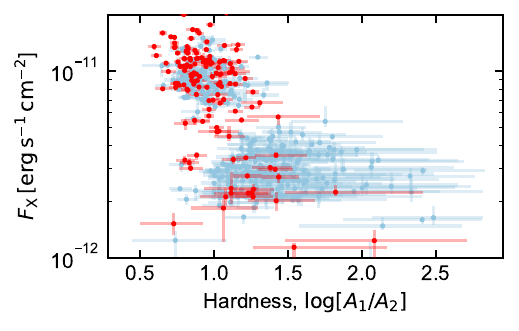}
    \caption{0.2--2~keV observed flux, $F_{\mathrm{X}}$ against an estimate of the spectral hardness ($\log [A_1 / A_2]$) during the early time NICER observations presented in Fig.~\ref{fig:nicer_spectral_evolution}. When $F_{\mathrm{X}}$ is higher, the spectrum becomes harder. The red markers denote observations obtained when the source exhibited large amplitude modulations during W1. A smaller $\log [A_1 / A_2]$ value denotes a harder X-ray spectrum.}
    \label{fig:nicer_spectral_evolution_simple}
\end{figure}

\begin{figure*}
    \centering
    \includegraphics[scale=0.59]{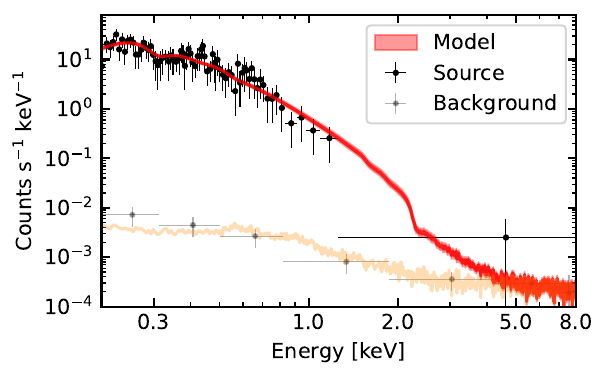}
    \includegraphics[scale=0.59]{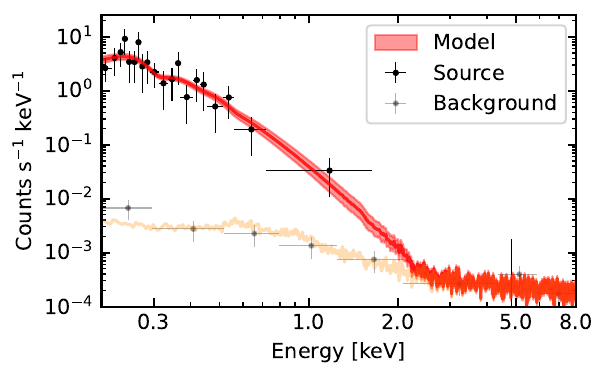}
    \includegraphics[scale=0.59]{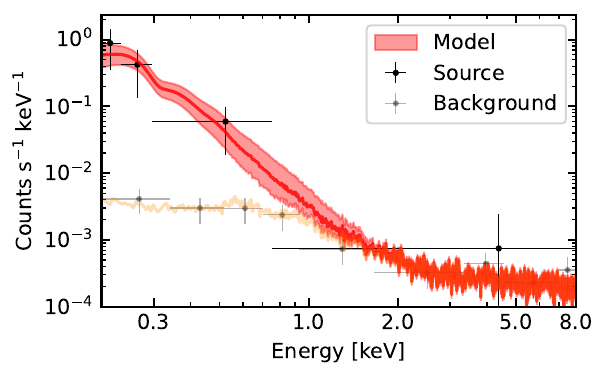}
    \includegraphics[scale=0.59]{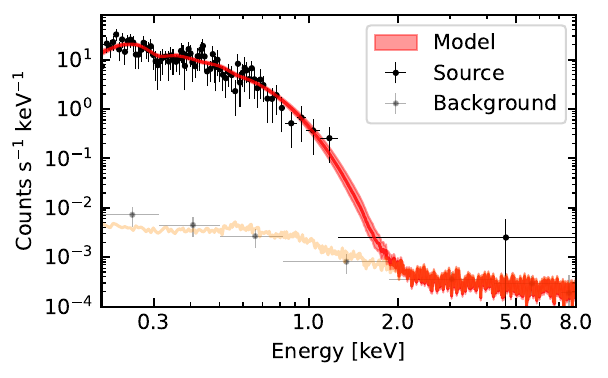}
    \includegraphics[scale=0.59]{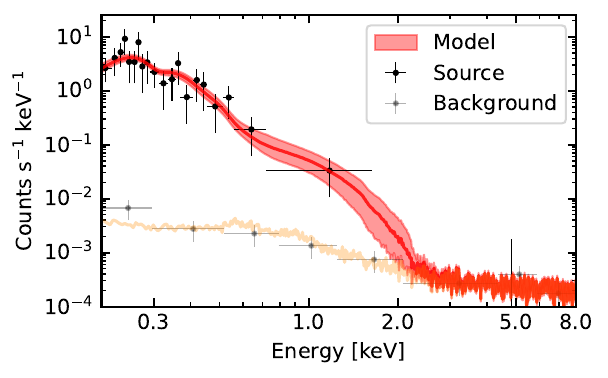}
    \includegraphics[scale=0.59]{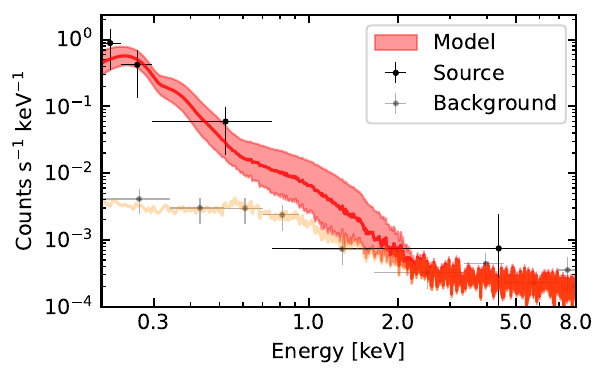}
    \caption{Power-law (\texttt{tbabs*zpowerlaw}, top row) and double blackbody (\texttt{tbabs*(zbbody+zbbody)}, bottom row) model fits to the convolved eROSITA spectra (eRASS2, left column; eRASS3, middle; eRASS4, right), with \namesrc \, being observed in eRASS2, eRASS3 and eRASS4 at 10, 192 and 378~days after optical peak. 
    The black and grey markers represent source and scaled background spectra. The solid red line denotes the median model, whilst the shaded red band encloses 68\,\% of the posterior, and the gold line denotes the median background model fit.
    Spectra are re-binned to minimum 10 counts per bin for plotting purpose only. The preferred model fit for the eRASS2 spectrum is the double blackbody model, whereas the power-law model is preferred for the eRASS3 and eRASS4 spectra. 
    }
    \label{fig:bxa_erosita_spectra}
\end{figure*}

\begin{figure*}
    \centering
    \includegraphics[scale=0.6]{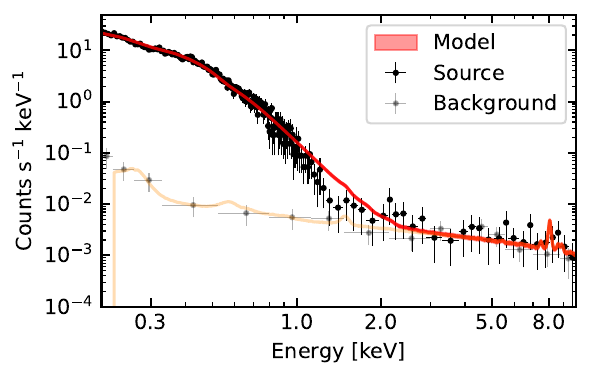}
    \includegraphics[scale=0.6]{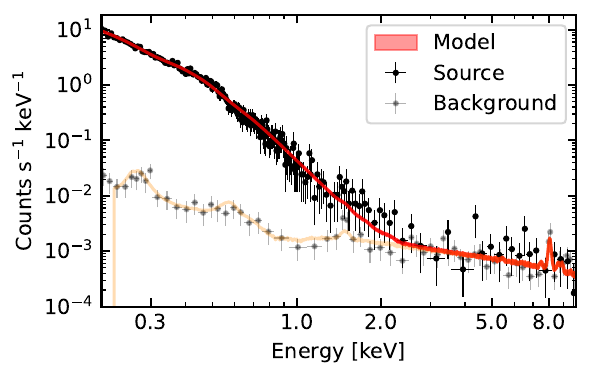}
    \includegraphics[scale=0.6]{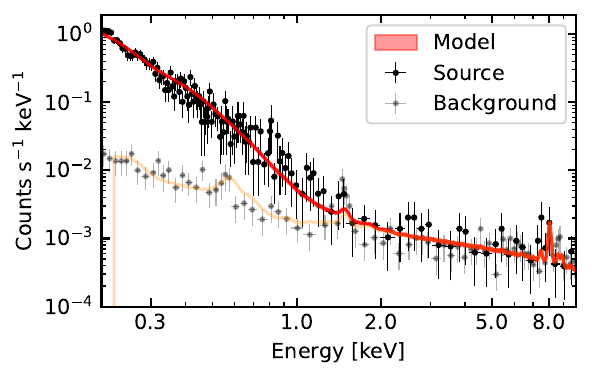}
    \includegraphics[scale=0.6]{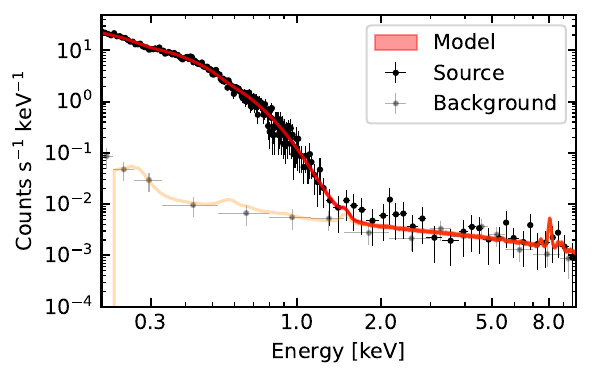}
    \includegraphics[scale=0.6]{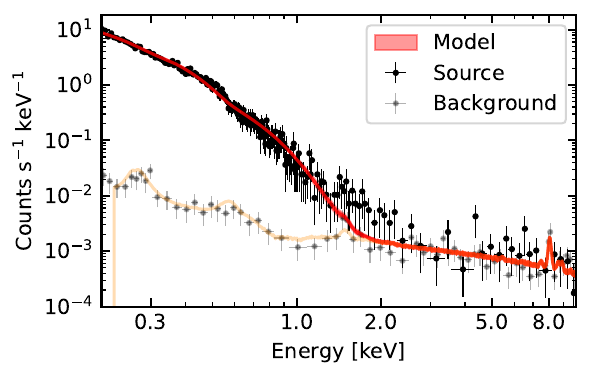}
    \includegraphics[scale=0.6]{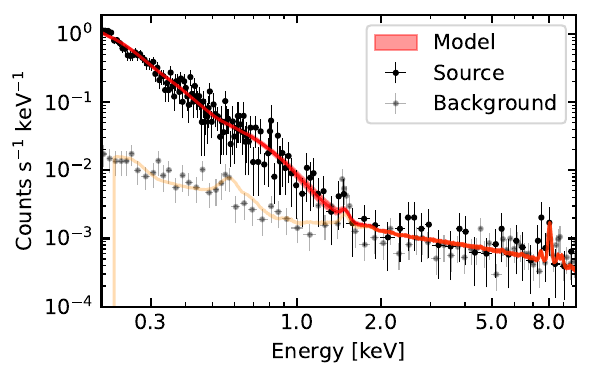}
    \caption{Absorbed power-law (\texttt{tbabs*ztabs*zpowerlaw}, top row) and double blackbody (\texttt{tbabs*(zbbody+zbbody)}, bottom row) model fits to the convolved XMM spectra (XMM1, left column; XMM2, middle; XMM3, right). The preferred model fit for XMM1 and XMM3 is the dual black body model, whereas it is the absorbed power-law model for XMM2. XMM1, XMM2 and XMM3 were obtained at 36, 196, and 393 days after optical peak.} 
    \label{fig:bxa_xmm_spectra}
\end{figure*}

The long-term evolution of the 0.2--2~keV flux and the photon index inferred from fitting a single power law to the X-ray spectra are shown in Fig.~\ref{fig:xray_photon_index_evolution}. 
The complete set of models considered for fitting to the X-ray spectra is presented in Table~\ref{tab:bxa_model_table}. Additional models—such as Comptonised disc emission or disc emission modified by ionised absorption—were omitted, as \citet{homan_discovery_2023} already previously investigated these possibilities for the XMM-Newton data of J2344 and found no significant improvement in the fit. The BXA spectra fit results for the eROSITA and XMM spectra can be found in Table~\ref{tab:x_ray_model_fits}; the best fitting spectral models to the eROSITA and XMM spectra are shown in Figs.~\ref{fig:bxa_erosita_spectra} and~\ref{fig:bxa_xmm_spectra}, respectively. A plot of the 0.2--2~keV flux against hardness for NICER observations can be found in Fig.~\ref{fig:nicer_spectral_evolution_simple}.

\begin{table}[h]
    \centering
    \caption{Spectral models for the eROSITA and XMM spectra.}
    \begin{tabular}{ll}
       Model  & Description \\
       \hline
       \texttt{zbbody}  & Redshifted blackbody \\
       \texttt{zpowerlaw}  & Redshifted power law \\
       \texttt{zbremss}  & Redshifted bremsstrahlung \\
       \texttt{zdiskbb}  & Redshifted accretion disc \\
       \texttt{ztbabs*zbbody}  & Redshifted blackbody,\\
       & with neutral host absorption \\
       \texttt{ztbabs*zpowerlaw}  & Redshifted power law,\\
       & with neutral host absorption \\
       \texttt{zbbody+zpowerlaw}  & Redshifted blackbody and power law \\
       \texttt{zbbody+zbbody}  & Redshifted dual blackbody \\
    \end{tabular}
    \label{tab:bxa_model_table}
\end{table}

\subsection{Significance of the NICER periodicity}\label{sec:appendix_periodicity}
\begin{figure}
    \centering
    \includegraphics[scale=0.8]{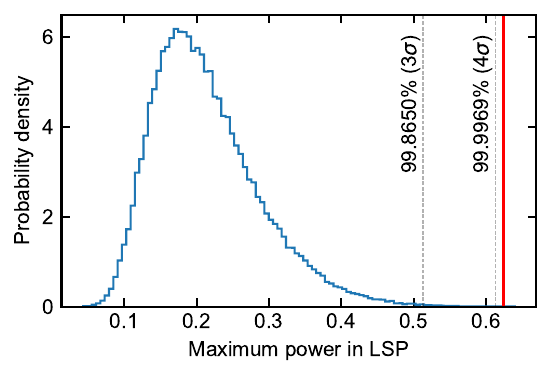}
    \caption{Distribution of maximum observed powers in the LSPs computed from each synthetic light curve. The solid red line marks the maximum power in the LSP computed from the observed 0.2--2 keV light curve (Fig.~\ref{fig:lsp_observed}), whilst the dashed grey lines are as defined in Fig.~\ref{fig:lsp_observed}.}
    \label{fig:max_power_histo}
\end{figure}

The LSP (\citealt{lomb_least-squares_1976,scargle_studies_1982}) computed using a least-squares normalisation \citep{vanderplas_understanding_2018} and from the high cadence X-ray observations obtained between MJD 59215 and 59240 (Fig.~\ref{fig:lsp_observed}) shows a clear peak at \tprec ~ days, with the uncertainty computed based on the FWHM of this peak. To further assess the statistical significance of this peak, we used the following framework:

\begin{enumerate}
    \item The 0.2--2 keV X-ray light curve was first fitted with a Gaussian process using the \texttt{celerite} package \citep{foreman-mackey_fast_2017}, and as in \citet{burke_characteristic_2021}, with a damped random walk (DRW) kernel of the form $ k(t_{nm}) =2\sigma^2 \exp{(-t_{nm}/\tau_{\rm DRW})} $, where $t_{nm}$ is the time separation between points $n$ and $m$ in the light curve, $\sigma$ is an amplitude decay term, and $\tau_{\rm DRW}$ the DRW timescale. From this Gaussian process regression, we infer $\log \sigma = -0.17 ^{+0.04}_{-0.03}$ and $\log [\tau _{\mathrm{DRW}}/ \, \mathrm{days}]=-0.24^{+0.16}_{-0.12}$. 
    \item $10^5$ synthetic light curves were generated from the power spectral density of the fitted Gaussian process using the method in \citet{timmer_generating_1995}, before modifying their probability distribution function to match the observed light curves as in \citet{emmanoulopoulos_generating_2013}, using the \texttt{lcsim} package \citep{kiehlmann_lcsim_2023}. These synthetic light curves were then down-sampled to match the sampling pattern of the observed light curves.
    \item For each synthetic light curve, we then computed its LSP using the same normalisation as above and measured its maximum power, $P_{\mathrm{max}}$.
    From the empirical distribution of $P_{\mathrm{max}}$ (Fig.~\ref{fig:max_power_histo}), then we compute the false-alarm probability of measuring $P \geq P_{\mathrm{max}}$, under the null hypothesis that there is no periodic component in the signal. 
    \item Only one of the synthetically generated light curves show a peak in their LSP as high as \OBSMAXPOWER \,(Fig.~\ref{fig:max_power_histo}), thus we consider the false-alarm probability to be $1 \times 10^{-5}$ (roughly equivalent to a $4.2 \sigma$ detection of periodicity).     
\end{enumerate}

However, we caveat this seemingly high significance detection with the issue that we have ultimately only observed a small number of putative cycles, due to the cutoff in NICER observations as the system entered sunblock. In addition, we caveat this with the fact that robust detections of quasi-periodicity in accreting SMBHs are notoriously challenging to obtain in cases where the light curves exhibit a low number of cycles. Lastly, our ability to model the continuum component of the power spectrum is severely hampered by the fact that the bulk of the light curve data available as input for \texttt{celerite} is dominated by the candidate quasi-periodic modulation signal. The DRW model parameters obtained here may be biased, as they describe a power spectrum containing only minimal red noise, and approaching white noise, over the bulk of the temporal frequencies probed. Consequently, these false-alarm probabilities are very likely biased downward.

\section{Photometric evolution}\label{sec:photometric_evolution} 

\begin{figure}[b]
    \centering
    \includegraphics[scale=0.65]{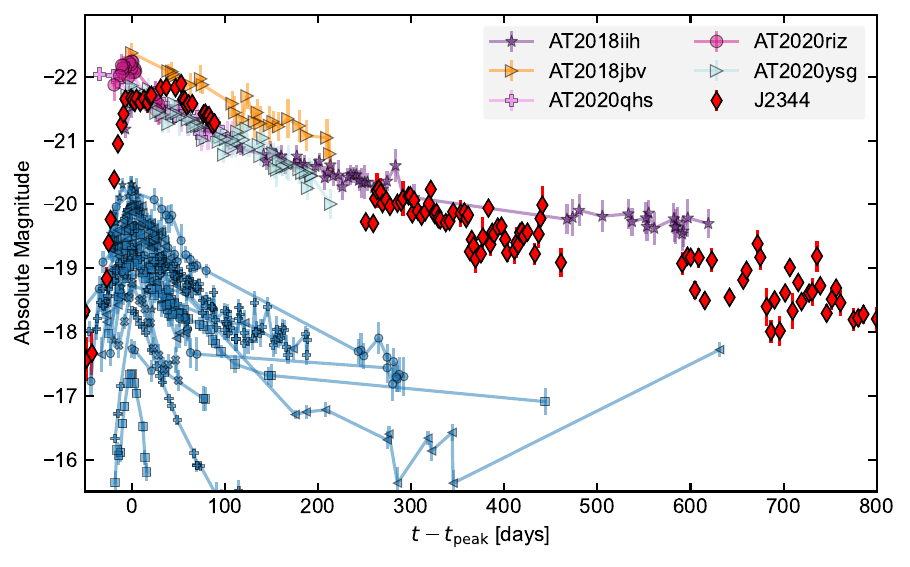}
    \caption{Comparison of the ATLAS $o$-band light curve of \namesrc \, with the ZTF $g$-band light curves of the sample of optically-selected TDEs (blue markers) in \citet{hammerstein_final_2023}.     }
    \label{fig:ztf_tde_lightcurve_comparison}
\end{figure}

The general rise and decay evolution of \namesrc \, in the optical, is broadly consistent with the observed behaviour of other optically-bright TDEs, with a comparison to a recent sample of ZTF-selected TDEs \citep{hammerstein_final_2023} presented in Fig.~\ref{fig:ztf_tde_lightcurve_comparison}. With a peak $c$-band absolute magnitude of $\sim$\absolutecmag, \namesrc \, is one of the most luminous observed TDE candidates identified to date, although this is still three mags fainter than the extreme accretion event AT2021lwx \citep{wiseman_multiwavelength_2023}. We note that there is a dearth of TDE candidates identified with peak absolute magnitudes approximately between -20~mag and -22~mag (Fig.~\ref{fig:ztf_tde_lightcurve_comparison}), and although such events are intrinsically rarer in the local universe \citep{van_velzen_mass_2018}, the reasons for such extreme luminosities are currently unclear.

\begin{table*}
\centering
\caption{X-ray spectral fit results for the best-fitting models shown in Figs. \ref{fig:bxa_erosita_spectra} and \ref{fig:bxa_xmm_spectra}.}
\label{tab:x_ray_model_fits}
\begin{tabular}{r|c|cc|ccc|cccccc}
\hline
Phase& ObsID & \multicolumn{2}{c|}{\texttt{zpowerlaw}} & \multicolumn{3}{c|}{\texttt{ztbabs*zpowerlaw}} & \multicolumn{6}{c}{\texttt{zbbody+zbbody}} \\
\hline
& & AIC & $\Gamma$ & AIC & $\log N_{\mathrm{H}}$ & $\Gamma$ & AIC & $kT_{\mathrm{BB1}}$ & $kT_{\mathrm{BB2}}$ & $\log A_{1}$ & $\log A_{2}$ & $\log A_{1}/A_{2}$ \\
 $[\rm{d}]$ & & & & & $ [\mathrm{cm}^{-2}]$ & & & [eV] & [eV] \\
\hline
10 & eRASS2 & 464 & $4.50 ^{+0.09}_{-0.09}$ & 460 & $20.06 ^{+0.19}_{-0.46}$ & $4.8 ^{+0.2}_{-0.2}$ & 454 & $43 ^{+5}_{-4}$ & $113 ^{+8}_{-6}$ & $-2.94 ^{+0.11}_{-0.10}$ & $-3.80 ^{+0.07}_{-0.10}$ & $0.88 ^{+0.09}_{-0.09}$ \\
36 & XMM1 & 3217 & $4.98 ^{+0.02}_{-0.02}$ & 2884 & $20.57 ^{+0.03}_{-0.03}$ & $5.9 ^{+0.1}_{-0.1}$ & 2811 & $51 ^{+1}_{-2}$ & $106 ^{+3}_{-3}$ & $-3.30 ^{+0.02}_{-0.02}$ & $-4.17 ^{+0.05}_{-0.05}$ & $0.87 ^{+0.04}_{-0.04}$ \\
192 & eRASS3 & 464 & $5.36 ^{+0.27}_{-0.24}$ & 465 & $19.23 ^{+0.79}_{-0.83}$ & $5.5 ^{+0.4}_{-0.3}$ & 469 & $50 ^{+4}_{-4}$ & $232 ^{+77}_{-69}$ & $-3.70 ^{+0.10}_{-0.09}$ & $-5.60 ^{+0.24}_{-0.25}$ & $1.89 ^{+0.21}_{-0.21}$ \\
196 & XMM2 & 5385 & $5.39 ^{+0.03}_{-0.03}$ & 5362 & $19.97 ^{+0.08}_{-0.11}$ & $5.7 ^{+0.1}_{-0.1}$ & 5480 & $49 ^{+1}_{-1}$ & $127 ^{+5}_{-5}$ & $-3.74 ^{+0.02}_{-0.02}$ & $-5.18 ^{+0.06}_{-0.05}$ & $1.44 ^{+0.04}_{-0.05}$ \\
378 & eRASS4 & 429 & $6.31 ^{+0.95}_{-0.88}$ & 431 & $19.21 ^{+0.81}_{-0.83}$ & $6.5 ^{+0.9}_{-0.9}$ & 431 & $39 ^{+8}_{-8}$ & $217 ^{+83}_{-108}$ & $-4.29 ^{+0.43}_{-0.33}$ & $-6.44 ^{+0.42}_{-0.52}$ & $2.25 ^{+0.44}_{-0.50}$ \\
393 & XMM3 & 5667 & $5.66 ^{+0.09}_{-0.09}$ & 5669 & $18.47 ^{+0.43}_{-0.33}$ & $5.7 ^{+0.1}_{-0.1}$ & 5647 & $40 ^{+1}_{-1}$ & $140 ^{+11}_{-9}$ & $-4.41 ^{+0.06}_{-0.06}$ & $-6.16 ^{+0.07}_{-0.07}$ & $1.75 ^{+0.05}_{-0.05}$ \\
\hline
\end{tabular}
\tablefoot{The Akaike information criterion (AIC) provides the goodness of fit. $\log A_1$ and $\log A_2$ are the normalisation of the softer (BB1) and harder (BB2) blackbody components for the \texttt{zbbody+zbbody} model. $\log(A_1/A_2)$ represents the spectral hardness. }
\end{table*}

% --------------------------

\end{appendix}

\end{document}